\documentstyle[prl,twocolumn,aps]{revtex}

\begin{document}
\draft

\title{Einstein-Podolsky-Rosen paradox without entanglement}

\author{Ryszard Horodecki$^{1,}$\cite{poczta1},
Micha\l{} Horodecki$^{1,}$\cite{poczta2} and Pawe\l{} Horodecki
$^{2,}$\cite{poczta3}}

\address{$^1$ Institute of Theoretical Physics and Astrophysics,
University of Gda\'nsk, 80--952 Gda\'nsk, Poland,\\
$^2$Faculty of Applied Physics and Mathematics,
Technical University of Gda\'nsk, 80--952 Gda\'nsk, Poland
}

\maketitle

\begin{abstract}
We claim that the  nonlocality without entanglement revealed quite
recently by Bennett {\it et al.} [quant-ph/9804053] should be
rather interpreted as
{\it Einstein-Podolsky-Rosen
paradox without entanglement}.
%It would have the status of the counterpart
It would be true nonlocality without entanglement
if one knew that quantum mechnics provides the best possible means for
extracting information from physical system
i.e. that it is ``operationally complete''.
\end{abstract}
\pacs{Pacs Numbers: 03.65.Bz}

Quite recently a very impressive effect called ``nonlocality without
entanglement''
has been revealed \cite{Bennett98} within information-theoretic framework.
The name ``nonlocality'' suggests that one deals with
version of the {\it Bell theorem} \cite{Bell} without entanglement. The purpose of
this note is to point out that  the effect should be instead treated as
 a counterpart of the {\it Einstein, Podolsky, and Rosen}  (EPR)
{\it paradox} \cite{EPR}.

Consider first the original EPR and Bell arguments. Without going into
details, one can say that EPR showed that if predictions of quantum
mechanics (QM) are correct then there is a conflict between two statements:

i) Nature can be described in a local-realistic way.

ii) Quantum mechanics provides a complete description of nature.

Call the view i) local realism (LR) (its negation is roughly
called nonlocality, see \cite{Stapp} in this context) and
ii) completness of quantum mechanics (CQM).
Then the EPR paradox can be written the form
\begin{equation}
QM\  {\rm predictions}\  \Rightarrow \ \sim (LR \wedge CQM).
\label{EPR}
\end{equation}

EPR believed that LR is true, so they concluded that QM is incomplete.
%Of course all it was done at believe that
It was Bell, who showed (by means of his  famous inequalities \cite{Bell})
that this conclusion is wrong: he found that
if we accept QM predictions, LR does not apply. So Bell argument
has the form
\begin{equation}
QM \ {\rm predictions} \ \Rightarrow \ \sim LR
\label{Bell}
\end{equation}
We see that Bell argument is apparently stronger statement than the EPR paradox.
This is compatible with the fact that Bell took into account more predictions
of quantum mechanics than EPR: the latter authors needed only
correlations coming from the same observables for both subsystems
(they used momentum and position),
while Bell employed also correlations between different observables.

Note that here one is able to verify where nonlocality enters the scene:
both EPR and Bell used predictions of QM for {\it entangled} states.
So, one can imagine that the pair of particles aquire nonlocal properties
in result of coupling which produced the entanglement.

Let us now turn to the effect of Bennett {\it et al.} \cite{Bennett98}
where entanglement was not used, and try to find why could we say
about nonlocality in this case. The scheme is the following.
There are three parties: Alice, Bob and Charlie. Alice and Bob are distantly
located and are allowed to communicate classically. Charlie prepares one of
nine {\it product} and {\it orthogonal}
states of two spin-1 particles and send one particle to Alice and the other
one to Bob. Bennett {\it et al.} \cite{Bennett98} showed that for a 
certain set of Charlie
states Alice and Bob cannot recover the state. So if we imagine
that Charlie wanted to send to Alice and Bob one of nine messages
represented by the states, Alice and Bob cannot read the message.
However, since the states are orthogonal, if Alice and Bob were not spatially
separated, they could measure the state and read the message  without any
error \cite{Peres}.  So one can argue that the information carried by the
pairs is nonlocal: it cannot be read from local properties of the systems and
correlations between these properties. One can say that  ``information
theoretic local realism'' (ILR) is violated.
This is what would be the counterpart of the
Bell theorem. However,  careful reader noted that in fact
we did not refute conjunction $${QM \ \rm predictions}  \ \wedge ILR$$ as it should be
to obtain Bell type argument. Indeed,
we tacitly assumed, that QM provides the {\it best possible means for
extraction of
information  on Charlie preparation} from the Alice and Bob particles.
Translating it into the language parallel to that of
EPR, we could say we assumed ``operational completness of QM'' (OCQM).
Now we come to formula exactly analogous to (\ref{EPR}):
\begin{equation}
QM \ {\rm predictions} \ \Rightarrow \ \sim (ILR \wedge OCQM)
\label{IEPR}
\end{equation}
However, since it is demonstrated using predictions of QM only for
{\it product} states, there is even stronger reason for refuting OCQM rather than
ILR, than in the case of the original EPR paradox: here it is hard to imagine,
when and  where the nonlocality enters.

In conclusion, we have argued that ``nonlocality without entanglement'' should
be regarded as ``EPR paradox without entanglement'' which still awaits
for Bell argument without entanglement to become true nonlocality without
entanglement. The main point is that one does not know, how to show that
QM provides us with the best possible tools for extracting information
from the single system. To obtain the true counterpart of  Bell argument,
one should
provide a proof for that. A possible way out could be the following:
better way of extraction of information will probably cause
violation of the no-cloning theorem \cite{nocloning}. This
violation would produce superluminal communication via entangled pair
of particles \cite{nocloning}.
In this way we would obtain that operational incompletness reduces
to sending superluminal signals, so we could replace CQM from the sentence
(\ref{IEPR}) with  statement that superluminal singnaling is impossible.
The problem is that we did it using entanglement... So, it seems that
it will be rather difficult  to produce Bell argument without
entanglement. Anyway, the Bennett {\it  et al.} effect has in fact the status
of EPR paradox without entanglement.

The authors would like to thank Marek \.Zukowski for numerous discussions on
EPR paradox and Bell's theorem.
M. H. and P. H. gratefully acknowledge the support from the Foundation for
Polish Science.

\end{document}